\documentclass[12pt]{article}
\usepackage{amssymb,amsmath,epsfig}
\allowdisplaybreaks

\begin{document}
\title{\textbf{Complexity Factor for Charged Spherical System}}

\author{M. Sharif \thanks{msharif.math@pu.edu.pk}~ and
Iqra Ijaz Butt \thanks{iqra.butt67@gmail.com}\\
Department of Mathematics, University of the Punjab,\\
Quaid-e-Azam Campus, Lahore-54590, Pakistan.}
\date{}
\maketitle

\begin{abstract}
In this paper, we study the complexity factor for a charged
anisotropic self-gravitating object. We formulate the
Einstein-Maxwell field equations, Tolman-Opphenheimer-Volkoff
equation, and the mass function. We form the structure scalars by
the orthogonal splitting of the Riemann tensor and then find the
complexity factor with the help of these scalars. Finally, we
investigate some astrophysical objects for the vanishing of
complexity condition. It is found that the presence of the
electromagnetic field decreases the complexity of the system.
\end{abstract}
\textbf{Keywords:} Self-gravitating system; Electromagnetic field;
Anisotropy; Complexity factor.\\
\textbf{PACS:} 04.40.-b; 04.40.Dg; 52.40.Db.

\section{Introduction}

The word complexity refers to a factor that includes all the terms
inducing complications in a system. Many attempts have been
dedicated towards a precise definition of complexity in various
sectors of science \cite{1}-\cite{11}. However, an exact definition
of complexity has not been obtained till now which defines it in
every field accumulatively. It may be noted that the definition of
complexity depends on the work established by Lopez-Ruiz and
collaborators \cite{7}-\cite{9}. In view of different proposed
definitions of complexity, it is related to the idea of information
and entropy that describe the structure of the system.

In physics \cite{7}, the term complexity begins by examining the
perfect crystal which has a periodic behavior and the isolated ideal
gas with a random behavior. A perfect crystal is a completely
ordered system of atoms that are arranged in a symmetric manner. A
small segment of information is sufficient to define the perfect
crystal which gives minimum complexity in the crystal. On the other
hand, the isolated gas is totally disordered and all the segments
have equal participation to give information related to the ideal
gas such that it has a maximum complexity. These systems are
examples of elementary models with extreme complexity.

The definition of complexity should also include some other factors
beyond information or order. Lopez-Ruiz et al. \cite{7} proposed the
abstract idea of disequilibrium which determines distance of the
probable distribution in the system of accessible state. Thus
disequilibrium should be maximum in case of perfect crystal and it
should be zero for the ideal gas. Consequently, disequilibrium and
information are introduced to describe the complexity by a quantity
that is a result of these two notions.

The definition of disequilibrium and information which include
probability distribution is redefined in \cite{12}-\cite{17} by the
term energy density in the fluid distribution. However, the term
energy density is not enough to describe the phenomenon of
complexity because the pressure is absent which appears in the
energy-momentum tensor and plays a vital role in the structure
formation of the fluid distribution.

In literature \cite{12}-\cite{17}, the idea of complexity has also
been applied to the self-gravitating systems like neutron stars and
white dwarfs. Recently, Herrera \cite{19} introduced a quite
different definition of complexity for a self-gravitating system.
This definition is related to the notion of structure of the
spherical system but is not related to disequilibrium or
information. He used the notion of Tolman mass which may be
considered the active gravitational mass for the fluid distribution.
This mass depends on the inhomogeneity of the energy density along
with anisotropy of the pressure. These two terms represent a single
scalar function which is a complexity factor. This vanishes when the
pressure is isotropic and energy density is homogenous and may also
vanish when the two notions namely, inhomogeneous energy density and
anisotropic pressure cancel each other. The variable which is
responsible for a complexity factor appears in the structure scalars
obtained from the orthogonal splitting of the Riemann tensor.

In literature, the study of charge in spherically symmetric
self-gravitating system started with the pioneer work of Rosseland
and Eddington \cite{20}. Bonnor \cite{22} examined the impact of
charge on spherical collapse of dust cloud and concluded that the
process of collapse slows down due to electric repulsion. Ray et al.
\cite{24} investigated the role of charge on compact stars and found
that $10^{20}$ Coulomb charge is present in the astrophysical
objects producing an electric field of $10^{21}V/m$. Sharif and
Bhatti \cite{24a} studied the effect of charge on the instability of
isotropic cylinder and deduced that the charge with other matter
variables control the stable as well as unstable configuration. The
same authors \cite{24b} investigated instability of charged
spherical system with the viscous dissipative matter distribution
and found the instability range from adiabatic index. Sharif and
Sadiq \cite{24c} analyzed the effect of electromagnetic field on the
stability of stellar object and concluded that it is stable for
specific choice of polytropic index. The same authors \cite{24d}
obtained exact solutions for anisotropic spherical system in the
presence of electromagnetic field and found that stability increases
with the effect of charge. Takisa and Maharaj \cite{25} studied the
charged anisotropic stellar solutions obeying polytropic equation of
state and found that the behavior of energy density and pressure are
consistent with the literature.

This paper studies the effects of charge on the definition of
complexity proposed by Herrera \cite{19}. The paper has the
following format. In the next section, we formulate basic equations
defining the structure of a stellar configuration. Section
\textbf{3} gives brief review of the orthogonal splitting of the
Riemann tensor as well as structure scalars. In section \textbf{4},
we introduce the complexity factor and obtain solutions of the
Einstein-Maxwell field equations for vanishing complexity factor.
Finally, we summarize our results in the last section.

\section{Basic Equations}

Here we discuss physical variables as well as the equations
necessary to define static charged stellar structure consisting of
anisotropic fluid. We consider static spherically symmetric geometry
in the interior of stellar structure defined by the line element
\begin{equation}\label{1}
ds^{2}=e^{\alpha(r)}dt^{2}-e^{\gamma(r)}dr^{2}-r^{2}(d\theta^{2}
+\sin^{2}\theta d\phi^{2}),
\end{equation}
bounded by the hypersurface $\Sigma$. We consider the
energy-momentum tensor for anisotropic fluid distribution as
\begin{equation}\label{2}
T^{\beta}_{\alpha} = \mu u^{\beta}u_{\alpha}-Ph^{\beta}_{\alpha}+\Pi
^{\beta}_{\alpha}+E ^{\beta}_{\alpha},
\end{equation}
where $\mu$ is the energy density and
\begin{equation}\label{3}
\Pi ^{\beta}_{\alpha} =
\Pi(s^{\beta}s_{\alpha}+\frac{1}{3}h^{\beta}_{\alpha}), \quad P =
\frac{1}{3}(P_{r}+ 2P_{\bot}), \quad \Pi = -(P_{\bot}-P_{r}), \quad
h^{\beta}_{\alpha} = \delta^{\beta}_{\alpha}-u^{\beta}u_{\alpha}.
\end{equation}
The four velocity and four-vector are defined by
\begin{equation}\label{4}
u^{\beta} = \bigg(\frac{1}{e^{\frac{\alpha}{2}}},0,0,0\bigg),\quad
s^{\beta} = \bigg(0,\frac{1}{e^{\frac{\gamma}{2}}},0,0\bigg),
\end{equation}
with the properties
\begin{equation}\label{5}
u^{\beta}u_{\beta}=1,\quad s^{\beta}s_{\beta}=-1, \quad
s^{\beta}u_{\beta}=0.
\end{equation}
The electromagnetic field tensor is given by
\begin{equation}\label{6}
E^{\beta}_{\alpha} = \frac{1}{4\pi}\bigg (-F^{\mu}_{\alpha}
F^{\beta}_{\mu}+\frac{1}{4} F^{\mu\nu} F_{\mu\nu}
\delta^{\beta}_{\alpha}\bigg),
\end{equation}
where $F_{\beta\alpha}$ is the Maxwell field tensor described by
$F_{\beta\alpha} = \phi_{\alpha,\beta}-\phi_{\beta,\alpha}$ and
$\phi_{\alpha}$ is the four potential determined by $\phi_{\beta} =
\phi\delta^{0}_{\beta}$.

The Maxwell field equations in four-vector formalism are given by
\begin{equation}\nonumber
F^{\beta\alpha}_{;\alpha} = \mu_{0} J^{\beta}, \quad
F_{[\beta\alpha;\gamma]}=0,
\end{equation}
where $\mu_{0}$ is the magnetic permeability and $J^{\beta}$ is the
four current defined by $J_{\beta} = \xi u_{\beta}$, where $\xi$ is
the charge density. For the metric (\ref{1}), the Maxwell field
equations yield
\begin{equation}\nonumber
\phi^{''}+\bigg(\frac{2}{r}-\frac{\alpha^{'}}{2}-\frac{\gamma^{'}}{2}\bigg)\phi^{'}
= 4\pi \xi e^{\frac{\alpha}{2}+\gamma}.
\end{equation}
Integration of the above equation gives
\begin{equation}\nonumber
\phi^{'}=\frac{e^\frac{\alpha+\gamma}{2}q(r)}{r^{2}},
\end{equation}
where
\begin{equation}\nonumber
q(r)=4\pi \int ^{r}_{0}\xi e^{\frac{\gamma}{2}}\hat{r}^{2}d\hat{r},
\end{equation}
represents the total charge within the sphere. The Einstein-Maxwell
field equations are
\begin{equation}\label{7}
G^{\alpha}_{\beta}= 8\pi (T^{\alpha}_{\beta} + E^{\alpha}_{\beta}),
\end{equation}
leading to
\begin{eqnarray}\label{8}
\mu&=&-\frac{1}{8\pi}\bigg[e^{-\gamma}\bigg(\frac{1}{r^{2}}-\frac{\gamma'}{r}\bigg)
-\frac{1}{r^{2}}+\frac{q^{2}}{r^{4}}\bigg],
\\\label{9}
P_{r}&=&-\frac{1}{8\pi}\bigg[\frac{1}{r^{2}}-e^{-\gamma}\bigg(\frac{1}{r^{2}}
+\frac{\alpha^{'}}{r}\bigg)-\frac{q^{2}}{r^{4}}\bigg],
\\\label{10}
P_{\bot}&=&\frac{e^{-\gamma}}{32\pi}\bigg(2\alpha^{''}+\alpha^{'2}
+2\frac{\alpha^{'}-\gamma^{'}}{r}-\gamma^{'}\alpha^{'}\bigg)-\frac{q^{2}}{8\pi
r^{4}},
\end{eqnarray}
where prime represents derivative with respect to $r$. The
conservation law gives the hydrostatic equilibrium equation
\begin{equation}\label{11}
P'_{r}=-\frac{\alpha'}{2}(\mu+P_{r})+\frac{2}{r}\bigg[(P_{\bot}
-P_{r})+\frac{qq'}{8\pi r^{3}}\bigg].
\end{equation}
This is also called the generalized Tolman-Opphenheimer-Volkoff
(TOV) equation for anisotropic charged fluid distribution. We
consider the Reissner-Nordstr\"{o}m metric for the exterior geometry
defined by
\begin{equation}\label{12}
ds^{2}= \bigg(1-
\frac{2M}{r}+\frac{Q^{2}}{r^{2}}\bigg)dt^{2}-\frac{dr^{2}}{\bigg(1
-\frac{2M}{r}+\frac{Q^{2}}{r^{2}}\bigg)}-r^{2}(d\theta^{2}+\sin^{2}\theta
d\phi^{2}),
\end{equation}
where $M$ and $Q$ denote the total mass and total charge in the
exterior region, respectively. The smooth matching of exterior and
interior spacetimes yields
\begin{equation}\label{13}
e^{\alpha}\overset\Sigma=1-\frac{2M}{r}+\frac{Q^{2}}{r^{2}}, \quad
e^{-\gamma}\overset\Sigma=1-\frac{2M}{r}+\frac{Q^{2}}{r^{2}},\quad
P_{r}\overset\Sigma=0.
\end{equation}
These equations are the necessary and sufficient conditions for
matching of the two metrics (\ref{1}) and (\ref{12}) on hypersurface
$\Sigma$.

Now, we evaluate the mass function using two definition namely,
Misner-Sharp mass and Tolman mass. The Misner-Sharp formula of mass
function \cite{29} yields
\begin{equation}\label{14}
m(r)=\frac{r}{2}(1-e^{-\gamma})+\frac{q^{2}}{2r}.
\end{equation}
Differentiation and then integration of Eq.(\ref{14}) and using
Eq.(\ref{8}), it follows that
\begin{equation}\label{15}
m(r)=4\pi\int^{r}_{0} \hat{r}^{2}\mu
d\hat{r}+\int^{r}_{0}\frac{qq'}{\hat{r}}d\hat{r}.
\end{equation}
Using the field equations and Eq.(\ref{14}), we obtain
\begin{eqnarray}\nonumber
m&=&\frac{4\pi}{3}r^{3}(\mu-P_{r}+P_{\bot})-\frac{r^{3}}{3}
\bigg[\frac{1}{4}e^{-\gamma}\bigg(\alpha^{''}
+\frac{\alpha^{'2}}{2}+\frac{\gamma^{'}}{r}
+\frac{2}{r^{2}}-\frac{\gamma^{'}\alpha^{'}}{2}\\\label{16}
&-&\frac{\alpha^{'}}{r}-\frac{2e^{\gamma}}{r^{2}}\bigg)
\bigg]+\frac{8\pi q^{2}}{3r}.
\end{eqnarray}
We simplify this expression using the Weyl tensor. The Weyl tensor
consists of two parts one is the magnetic part which vanishes for
spherical system while the other is the electric part defined as
\begin{equation}\label{17}
E_{\alpha\beta} = C_{\alpha\gamma\beta\delta}u^{\gamma}u^{\delta},
\end{equation}
where
\begin{equation}\label{18}
C_{\nu\mu\kappa\gamma} =
(g_{\nu\mu\alpha\beta}g_{\kappa\gamma\rho\delta}-\eta_{\nu\mu\alpha\beta}
\eta_{\kappa\gamma\rho\delta})u^{\mu}u^{\rho}E^{\nu\delta},
\end{equation}
with
$g_{\beta\alpha\mu\nu}=g_{\beta\mu}g_{\alpha\nu}-g_{\beta\nu}g_{\alpha\mu}$
and $\eta_{\beta\alpha\mu\nu}$ represents the Levi-Civita tensor.
$E_{\alpha\beta}$ is defined as
\begin{equation}\label{19}
E_{\alpha\beta} =
\mathcal{E}\left(\frac{1}{3}h_{\alpha\beta}+s_{\alpha}s_{\beta}\right),
\end{equation}
with
\begin{equation}\label{20}
\mathcal{E}=
-\frac{1}{4}e^{-\gamma}\bigg[\alpha^{''}+\frac{\alpha^{'2}}{2}
+\frac{\gamma^{'}}{r} +\frac{2}{r^{2}}
-\frac{\gamma^{'}\alpha^{'}}{2}
-\frac{\alpha^{'}}{r}-\frac{2e^{\gamma}}{r^{2}}\bigg].
\end{equation}
Using Eq.(\ref{20}) in (\ref{16}), we have
\begin{equation}\label{21}
m=\frac{4\pi}{3}r^{3}(\mu-P_{r}+P_{\bot})+\frac{1}{3}r^{3}\mathcal{E}
+\frac{8\pi q^{2}}{3r}.
\end{equation}
Comparing Eqs.(\ref{15}) and (\ref{21}), it follows that
\begin{equation}\label{22}
\mathcal{E}=4\pi(P_{r}-P_{\bot})-\frac{4\pi}{r^{3}}\int^{r}_{0}
\hat{r}^{3}\mu^{'}d\hat{r}-8\pi \frac{q^{2}}{r^{4}}+\frac{3}{r^{3}}
\int^{r}_{0}\frac{qq^{'}}{\hat{r}}d\hat{r}.
\end{equation}
Substituting the above equation in (\ref{21}), we obtain
\begin{equation}\label{23}
m=\frac{4\pi
r^{3}}{3}\mu-\frac{4\pi}{3}\int^{r}_{0}\hat{r}^{3}\mu^{'}d\hat{r}
+\int^{r}_{0}\frac{qq^{'}}{\hat{r}}d\hat{r}.
\end{equation}
Equation (\ref{22}) expresses $\mathcal{E}$ in terms of physical
quantities namely, inhomogeneous density, anisotropic pressure as
well as total charge and Eq.(\ref{23}) represents the corresponding
expression of mass function. Using Eq.(\ref{14}) in (\ref{9}), it
follows that
\begin{equation}\label{24}
\alpha' = 2 \quad \frac{4\pi r^{4}
P_{r}+rm-q^{2}}{r(r^{2}+q^{2}-2rm)}.
\end{equation}
Inserting Eq.(\ref{24}) in Eq.(\ref{11}), we obtain the following
form of TOV equation
\begin{equation}\label{25}
P'_{r}=-\frac{4\pi r^{4}
P_{r}+rm-q^{2}}{r(r^{2}+q^{2}-2rm)}(\mu+P_{r})+\frac{2}{r}
\bigg[(P_{\bot}-P_{r})+\frac{qq'}{8\pi r^{3}}\bigg].
\end{equation}

Tolman \cite{30} proposed another definition of energy for static
spherical system defined by
\begin{equation}\label{26}
m_{T}=4\pi \int^{r_{\Sigma}}_{0}
r^{2}e^{\frac{\alpha+\gamma}{2}}(\mu+P_{r}+2P_{\bot})dr.
\end{equation}
The total energy of the fluid within the sphere of radius $r$ is
\begin{equation}\label{27}
m_{T}=4\pi \int^{r}_{0}
\hat{r}^{2}e^{\frac{\alpha+\gamma}{2}}(\mu+P_{r}+2P_{\bot})d\hat{r}.
\end{equation}
Using Eqs.(\ref{8})-(\ref{10}) in (\ref{27}), it follows that
\cite{31,32}
\begin{equation}\label{28}
m_{T}=\frac{r^{2}}{2}e^{\frac{\alpha-\gamma}{2}}\alpha^{'}-\int
^{r}_{0}
e^{\frac{\alpha+\gamma}{2}}\frac{q^{2}}{\hat{r}^{2}}d\hat{r}.
\end{equation}
Putting the value of $\alpha'$ from Eq.(\ref{24}) into (\ref{28}),
the Tolman mass becomes
\begin{equation}\label{29}
m_{T}=\frac{e^{\frac{\alpha+\gamma}{2}}(4\pi
r^{4}P_{r}+rm-q^{2})}{r}-\int ^{r}_{0}
e^{\frac{\alpha+\gamma}{2}}\frac{q^{2}}{\hat{r}^{2}}d\hat{r}.
\end{equation}
Equation (\ref{28}) can also be interpreted as active gravitational
mass of the system. Another expression for $m_{T}$ \cite{31,32} is
\begin{eqnarray}\nonumber
m_{T}&=&\bigg(\frac{r}{r_{\Sigma}}\bigg)^{3}[m_{T}]_{\Sigma}-r^{3}
\int^{r_{\Sigma}}_{r}\frac{e^{\frac{\alpha+\gamma}{2}}}{\hat{r}}[4\pi
(P_{\bot}-P_{r})-\mathcal{E}]d\hat{r}+r^{3}\int^{r_{\Sigma}}_{r}e^{\frac{\alpha
+\gamma}{2}}\frac{3q^{2}}{\hat{r}^{5}}d\hat{r}\\\label{30}
&-&r^{3}\int^{r_{\Sigma}}_{r}e^{\frac{\alpha+\gamma}{2}}\frac{8\pi
q^{2}}{\hat{r}^{5}}d\hat{r}.
\end{eqnarray}
Using Eq.(\ref{22}), this equation turns out to be
\begin{eqnarray}\nonumber
m_{T}&=&\bigg(\frac{r}{r_{\Sigma}}\bigg)^{3}[m_{T}]_{\Sigma}-r^{3}
\int^{r_{\Sigma}}_{r}e^{\frac{\alpha+\gamma}{2}}
\bigg[\frac{8\pi}{\hat{r}}(P_{\bot}-P_{r})+\frac{1}
{\hat{r}^{4}}\int^{\hat{r}}_{0}4\pi
\hat{r}^{3}\mu^{'}d\hat{r}\\\label{31}&+&(8\pi-\frac{3}{2})
\frac{q^{2}}{\hat{r}^{4}}\bigg]d\hat{r}+r^{3}
\int^{r_{\Sigma}}_{r}e^{\frac{\alpha+\gamma}{2}}
\frac{3q^{2}}{\hat{r}^{5}}d\hat{r}-r^{3}\int^{r_{\Sigma}}_{r}
e^{\frac{\alpha+\gamma}{2}}\frac{8\pi q^{2}}{\hat{r}^{5}}d\hat{r}.
\end{eqnarray}
This represents the contribution of density inhomogeneity, charge
and anisotropy of pressure in the Tolman mass.

\section{Structure Scalars}

In this section, we use orthogonal splitting of the Riemann tensor
introduced by Bel \cite{33} and obtain scalar structures which help
us to find the complexity factor. For the orthogonal splitting of
the Riemann tensor, the following tensors are introduced
\cite{33}-\cite{35}
\begin{eqnarray}\label{32}
Y_{\alpha\beta}&=&R_{\alpha\gamma\beta\delta}u^{\gamma}u^{\delta},
\\\label{33}
X_{\alpha\beta}
&=&^{*}R^{*}_{\alpha\gamma\beta\delta}u^{\gamma}u^{\delta}=\frac{1}{2}\eta
^{\epsilon\nu}_{\alpha\gamma}R^{*}_{\epsilon\nu\beta\delta}u^{\gamma}u^{\delta},
\end{eqnarray}
where $R^{*}_{\mu\nu\gamma\delta}=\frac{1}{2}\eta
_{\epsilon\beta\gamma\delta}R^{\epsilon\beta}_{\mu\nu}$. These
tensors can be expressed in the trace-free and trace parts as
\begin{eqnarray}\label{34}
Y_{\alpha\beta}&=&Y_{TF}\bigg(\frac{1}{3}h_{\alpha\beta}+s_{\alpha}s_{\beta}\bigg)
+\frac{1}{3}Y_{T}h_{\alpha\beta},
\\\label{35}
X_{\alpha\beta}&=&X_{TF}\bigg(\frac{1}{3}h_{\alpha\beta}+s_{\alpha}s_{\beta}\bigg)
+\frac{1}{3}X_{T}h_{\alpha\beta}.
\end{eqnarray}
Using the Einstein-Maxwell filed equations, we obtain these scalars
as \cite{35}
\begin{eqnarray}\label{36}
X_{T}&=&8\pi\mu+\frac{q^{2}}{r^{4}}, \\\label{37}
X_{TF}&=&4\pi\Pi+\frac{q^{2}}{r^{4}}-\mathcal{E}.
\end{eqnarray}
Also, using Eq.(\ref{22}), we have
\begin{equation}\label{38}
X_{TF}=\frac{4\pi}{r^{3}}\int^{r}_{0}\hat{r}^{3}\mu^{'}d\hat{r}
+(8\pi-\frac{1}{2})\frac{q^{2}}{r^{4}}.
\end{equation}

The expressions for $Y_{T}$ and $Y_{TF}$ are
\begin{eqnarray}\label{39}
Y_{T}&=&4\pi(\mu-2\Pi+3P_{r})+\frac{q^{2}}{r^{4}}, \\\label{40}
Y_{TF}&=&4\pi\Pi+\frac{q^{2}}{r^{4}}+\mathcal{E}.
\end{eqnarray}
Substitution of Eq.(\ref{22}) in (\ref{40}) yields
\begin{equation}\label{41}
Y_{TF}=(\frac{5}{2}-8\pi)\frac{q^{2}}{r^{4}}-\frac{4\pi}{r^{3}}
\int^{r}_{0}\hat{r}^{3}\mu^{'}d\hat{r}+8\pi\Pi.
\end{equation}
The scalars $X_{TF}$ and $Y_{TF}$ describe anisotropy of the
pressure in the presence of charge as
\begin{equation}\label{42}
Y_{TF}+X_{TF}=\frac{2q^{2}}{r^{4}}+8\pi\Pi.
\end{equation}
Replacing Eq.(\ref{40}) in (\ref{30}), we obtain
\begin{eqnarray}\nonumber
m_{T}&=&\bigg(\frac{r}{r_{\Sigma}}\bigg)^{3}[m_{T}]_{\Sigma}+r^{3}
\int^{r_{\Sigma}}_{r}\frac{e^{\frac{\alpha+\gamma}{2}}}{\hat{r}}
Y_{TF}d\hat{r} -r^{3}\int^{r_{\Sigma}}_{r}
e^{\frac{\alpha+\gamma}{2}}8\pi\frac{q^{2}}{\hat{r}^{5}}d\hat{r}
\\\label{43} &+&r^{3}\int^{r_{\Sigma}}_{r}e^{\frac{\alpha
+\gamma}{2}}\frac{2q^{2}}{\hat{r}^{5}}d\hat{r}.
\end{eqnarray}
Comparing Eq.(\ref{43}) with (\ref{30}), it follows that
\begin{eqnarray}\nonumber
r^{3}\int^{r_{\Sigma}}_{r}\frac{e^{\frac{\alpha+\gamma}{2}}}{\hat{r}}
Y_{TF}d\hat{r}&=&-r^{3}\int^{r_{\Sigma}}_{r}\frac{e^{\frac{\alpha
+\gamma}{2}}}{\hat{r}}[4\pi (P_{\bot}-P_{r})-\mathcal{E}]d\hat{r}
\\\nonumber&+&r^{3}\int^{r_{\Sigma}}_{r}e^{\frac{\alpha+\gamma}{2}}
\frac{3q^{2}}{\hat{r}^{5}}d\hat{r}
-r^{3}\int^{r_{\Sigma}}_{r}e^{\frac{\alpha
+\gamma}{2}}\frac{2q^{2}}{\hat{r}^{5}}d\hat{r}.
\end{eqnarray}
This shows that $Y_{ TF}$ is associated with the effect of the
anisotropic pressure, inhomogeneity of the energy density and total
charge of the fluid distribution, i.e., $Y_{TF}$ describes the
effect of these two quantities in the Tolman mass defined in
Eq.(\ref{43}). Also, the Tolman mass given in Eq.(\ref{26}) can be
written in terms of structure scalar as
\begin{equation}\label{44}
m_{T}=\int^{r}_{0}\hat{r}^{2}e^{\frac{\alpha+\gamma}{2}}
\bigg(Y_{T}-\frac{q^{2}}{\hat{r}^{4}}\bigg)d\hat{r}.
\end{equation}

\section{The Complexity Factor}

There are many factors producing complexity in a system for example,
density inhomogeneity, pressure anisotropy, electromagnetic field,
heat dissipation and viscosity. In general, any system having
homogenous energy density as well as isotropic pressure and in the
absence of the above mentioned factors is considered as the simplest
system with negligible complexity. For our fluid, the causes of
complexity are inhomogeneous energy density, anisotropic pressure
and electromagnetic field. The structure scalar $Y_{TF}$ defined in
Eq.(\ref{41}) contains these terms (inhomogeneous energy density,
anisotropic pressure and charge) which are responsible for producing
the complexity in the system. For this reason, the term complexity
factor can be associated with the structure scalar $Y_{TF}$.

Furthermore, the structure scalar $Y_{TF}$ appears in the Tolman
mass implying that these terms affect Tolman mass. Now, we discuss
the vanishing complexity condition. The set of field equations leads
to a system of three ordinary differential equations in which there
are five unknown functions $\alpha,\gamma,\mu,P_{r},P_{\bot}$. Using
the condition $Y_{TF}=0$, we are left with four unknowns and need
one more condition to have a unique solution. For this purpose, we
take Eq.(\ref{41}) and use $Y_{TF}=0$. The vanishing complexity
condition gives
\begin{equation}\label{45}
\Pi=\left(1-\frac{5}{16\pi}\right)\frac{q^{2}}{r^{4}}
+\frac{1}{2r^{3}}\int^{r}_{0}\hat{r}^{3}\mu^{'}d\hat{r}.
\end{equation}

In the following, we discuss some examples.

\subsection{The Gokhroo and Mehra Ansatz}

Gokhroo and Mehra \cite{36} discussed the internal structure of the
spherical configuration with the variable energy density for
anisotropic spheres. Physically, these solutions have been used to
discuss the behavior of compact objects. Here, we use the assumption
proposed by Gokhroo and Mehra to discuss the behavior of stellar
structures for vanishing complexity condition. The proposed energy
density is
\begin{equation}\label{46}
\mu=\mu_{0}\bigg(1-\frac{Kr^{2}}{r^{2}_{\Sigma}}\bigg),
\end{equation}
where $K=(0,1)$.
 Using this value in Eq.(\ref{15}), we have
\begin{equation}\label{47}
m(r)=\int^{r}_{0}\frac{qq^{'}}{\hat{r}}d\hat{r}-\frac{4\pi
r^{3}}{3}\mu_{0}\bigg(\frac{3Kr^{2}}{5r^{2}_{\Sigma}}-1\bigg).
\end{equation}
Inserting this value in Eq.(\ref{14}), we obtain
\begin{equation}\label{48}
e^{-\gamma}=1+\frac{3K\hat{\alpha}
r^{4}}{5r^{2}_{\Sigma}}+\frac{q^{2}}{r^{2}}-\hat{\alpha}
r^{2}-\frac{2}{r}\int^{r}_{0}\frac{qq^{'}}{\hat{r}}d\hat{r},
\end{equation}
where $\hat{\alpha}=8\pi\mu_{0}/3$. From Eqs.(\ref{9}) and
(\ref{10}), it follows that
\begin{equation}\label{49}
8\pi(P_{r}-P_{\bot})=e^{-\gamma}\bigg[\frac{\alpha^{'}}{2r}
+\frac{1}{r^{2}}+\frac{\gamma^{'}\alpha^{'}}{4}
+\frac{\gamma^{'}}{2r}-\frac{\alpha^{''}}{2}
-\frac{\alpha^{'2}}{4}\bigg]-\frac{1}{r^{2}}+\frac{2q^{2}}{r^{4}}.
\end{equation}

Introducing the new variables \cite{19}
\begin{equation}\label{50}
e^{\alpha(r)}=e^{-\int(\frac{2}{r}-2z(r))dr},\quad
1/e^{\gamma}=y(r).
\end{equation}
Using these variables, we obtain the following form of Eq.(\ref{49})
\begin{equation}\label{51}
y^{'}-y\bigg[\frac{6}{r}-\frac{2z^{'}}{z}-2z-\frac{4}{r^{2}z}\bigg]
+\frac{4q^{2}}{zr^{4}}=-\bigg(\frac{1}{r^{2}}+8\pi\Pi\bigg)\frac{2}{z}.
\end{equation}
Its solution leads to the line element in terms of $z$ and $\Pi$
\cite{37} as
\begin{eqnarray}\nonumber
ds^{2}&=&-e^{-\int(\frac{2}{r}-2z(r))dr}dt^{2}+r^{2}d\theta^{2}+r^{2}\sin^{2}\theta
d\phi^{2}\\\label{52} &+&\frac{z^{2}(r)e^{\int
(2z(r)+\frac{1}{r^{2}z(r)})dr}}{r^{6}(-2\int \frac{e^{\int
(2z(r)+\frac{1}{r^{2}z(r)})dr}z(r)(1+8\pi\Pi r^{2})}{r^{8}}dr -\int
\frac{4q^{2}}{zr^{4}}dr +C)}dr^{2},
\end{eqnarray}
where C is an integration constant. Using the value of $q(r)$ and
Eq.(\ref{46}), the vanishing complexity condition (\ref{45}) becomes
\begin{equation}\label{53}
\Pi=\bigg(1-\frac{5}{16\pi}\bigg)\frac{[4\pi\int^{r}_{0}\xi
(\frac{1}{y})^{\frac{1}{2}}\hat{r}^{2}d\hat{r}]^{2}}
{r^{4}}-\frac{\mu_{0}K}{r^{3}}\int^{r}_{0}\frac{\hat{r}^{4}}{\hat{r}^{2}_{\Sigma}}d\hat{r}.
\end{equation}
The physical variables in the presence of charge yield
\begin{eqnarray}\label{54}
P_{r}&=&\frac{\frac{m}{r}-z(2m-r)-1+\frac{q^{2}}{2r^{2}}}{4\pi
r^{2}},
\\\label{55}
\mu&=&\frac{m^{'}}{4\pi r^{2}}-\frac{qq^{'}}{4\pi r^{3}},
\\\label{56}
P_{\bot}&=&\frac{1}{8\pi}\bigg[z\bigg(\frac{m}{r^{2}}-\frac{m^{'}}{r}\bigg)
+\bigg(z^{'}+z^{2}-\frac{z}{r}+\frac{1}{r^{2}}\bigg)\bigg(1-\frac{2m}{r}\bigg)
-\frac{q^{2}}{r^{4}}\bigg].
\end{eqnarray}
Equations (\ref{54})-(\ref{56}) indicate the presence of charge in
radial pressure, energy density and tangential pressure,
respectively.

\subsection{Polytropic Equations with Vanishing Complexity Factor}

For self-gravitating system, the polytropic equation plays a vital
role. Polytropes with anisotropic matter distribution have widely
been discussed in literature \cite{38,39}. Here, we discuss two
different cases related to the polytropes. One of them is
\begin{equation}\label{57}
P_{r}=\mathcal{K}\mu^{\sigma}=\mathcal{K}\mu^{1+1/n},
\end{equation}
where $\mathcal{K}$ is called polytropic constant, $\sigma$ is
called polytropic exponent and $n$ is called polytropic index. We
take TOV equation (\ref{11}) and convert it into the dimensionless
form. For this purpose, we use dimensionless variables
\begin{eqnarray}\label{58}
\beta&=&\frac{P_{rc}}{\mu_{c}}, \quad r=\frac{\zeta}{A}, \quad
A^{2}=\frac{4\pi \mu_{c}}{\beta(n+1)}, \\\label{59}
v(\zeta)&=&\frac{A^{3}m(r)}{4\pi \mu_{c}}, \quad
\Psi^{n}=\frac{\mu}{\mu_{c}},
\end{eqnarray}
where subscript $c$ shows that the quantity is calculated at the
center. At the boundary $r=r_{\Sigma}$ ($\zeta=\zeta_{\Sigma}$), we
have $\Psi(\zeta_{\Sigma})=0$ \cite{38}. Putting the values from
Eqs.(\ref{24}) and (\ref{57})-(\ref{59}) in (\ref{11}), we obtain
\begin{eqnarray}\nonumber
&&\bigg[\frac{1-\frac{2v \beta(n+1)}{\zeta}+\frac{4\pi
q^{2}\mu_{c}}{\beta(n+1)\zeta^{2}}}{1+\beta\Psi}\bigg]
\bigg(\zeta^{2}\frac{d\Psi}{d\zeta}
+2\Psi^{-n}\frac{\beta^{4}(n+1)^{2}\zeta^{3}\Pi-2\pi
\beta^{2}\mu^{2}_{c}q
\frac{dq}{d\zeta}}{\beta^{5}\mu_{c}\zeta^{2}(n+1)^{3}}\bigg)\\\label{60}
&&+ \beta \zeta^{3} \Psi^{n+1}+v-\frac{4\pi
q^{2}\mu_{c}}{\zeta\beta^{2}(n+1)^{2}}=0.
\end{eqnarray}
We also convert Eq.(\ref{15}) in dimensionless variables as
\begin{equation}\label{61}
\frac{dv}{d\zeta}=\zeta^{2}\Psi^{n}+\frac{dq}{d\zeta}\frac{4\pi
q\mu_{c}}{\beta^{2}\zeta(n+1)^{2}}.
\end{equation}
Equations (\ref{60}) and (\ref{61}) are two ordinary differential
equations with three unknown functions $\Pi,v,\Psi$. We still need
one more condition to have a unique solution. For this purpose, we
use the vanishing complexity condition (\ref{45}) in dimensionless
form given by
\begin{equation}\label{62}
\frac{2\zeta}{n\mu_{c}}\frac{d\Pi}{d\zeta}+\frac{6\Pi}{n\mu_{c}}=
\Psi^{n-1}\zeta\frac{d\Psi}{d\zeta}+\bigg(1-\frac{5}{16\pi}\bigg)
\bigg(\frac{2qq'A^{3}}{\zeta^{4}}-\frac{4q^{2}A^{4}}{\zeta^{5}}\bigg).
\end{equation}

Now, we have three differential equations (\ref{60})-(\ref{62}) with
three unknown functions $\Pi,v,\Psi$. For any of value of $n$ and
$\beta$, this system can be integrated analytically or we can have a
numerical solution using appropriate initial conditions. These
equations physically describe the structure of stellar objects with
the vanishing complexity condition. Any solution of this system
gives the pressure, density, mass and radius of a specific stellar
object for the chosen values of free parameters. We discuss another
case of polytropes with the equation of state
$P_{r}=\mathcal{K}\mu^{\sigma}_{d}=\mathcal{K}\mu^{1+\frac{1}{n}}_{d},$
where $\mu_{d}$ represents the baryonic mass density. Following the
above procedure, we obtain
\begin{eqnarray}\nonumber
&&\bigg[\frac{1-\frac{2v \beta(n+1)}{\zeta}+\frac{4\pi
q^{2}\mu_{c}}{\beta(n+1)\zeta^{2}}}{1+\beta\Psi_{d}}\bigg]
\bigg(\zeta^{2}\frac{d\Psi_{d}}{d\zeta}
+2\Psi^{-n}_{d}\frac{\beta^{4}(n+1)^{2}\zeta^{3}\Pi-2\pi
\beta^{2}\mu^{2}_{c}q
\frac{dq}{d\zeta}}{\beta^{5}\mu_{c}\zeta^{2}(n+1)^{3}}\bigg)\\\label{64}&&+\beta
\zeta^{3} \Psi^{n+1}_{d}+v-\frac{4\pi
q^{2}\mu_{c}}{\zeta\beta^{2}(n+1)^{2}}=0,
\end{eqnarray}
\begin{eqnarray}\nonumber
\frac{2\zeta}{n\mu_{dc}}\frac{d\Pi}{d\zeta}+\frac{6\Pi}{n\mu_{dc}}
&=&\Psi^{n-1}_{d}\zeta\frac{d\Psi_{d}}{d\zeta}[\mathcal{K}(n+1)
\mu^{1/n}_{dc}\Psi_{d}+1]\\\label{65}&+&\bigg(1-\frac{5}{16\pi}\bigg)
\bigg(\frac{2qq'A^{3}}{\zeta^{4}}-\frac{4q^{2}A^{4}}{\zeta^{5}}\bigg),
\end{eqnarray}
with
\begin{equation}\nonumber
\Psi^{n}_{d}=\frac{\mu_{d}}{\mu_{dc}}.
\end{equation}
However, Eq.(\ref{61}) remains the same for this equation of state.
Again, we have a system of differential equations ((\ref{61}),
(\ref{64}), (\ref{65})) defining the structure of stellar
configuration with specific equation of state and zero complexity
factor.

\section{Conclusions}

In astrophysics, the study of stellar objects is an interesting
phenomenon due to their physical features which motivate the
researchers to explore these objects. Different aspects including
mass-radius ratio, luminosity, anisotropy and stability or
instability of stellar configurations have widely been studied in
literature. However, the term complexity factor is not studied in
detail for stellar objects. In this paper, we have studied the
complexity factor for charged spherically symmetric stellar objects
and the behavior of these objects in the context of vanishing
complexity condition. This provides the effects of electromagnetic
field on Herrera's work \cite{19}. We have formulated the
Einstein-Maxwell field equations and found the mass function using
Misner-Sharp as well as Tolman formalism. We have discussed
structure scalars in the presence of electromagnetic field and
obtained the complexity factor. The complexity factor $Y_{TF}$
(\ref{41}) contains the terms associated with energy density
inhomogeneity, charge and anisotropic pressure. This equation
indicates that the inclusion of charge decreases the complexity of
the system.

Moreover, using the assumption $Y_{TF}=0$ we have investigated the
vanishing complexity condition (defined in Eq.(\ref{45})) for two
examples of self-gravitating systems studied in the literature.
Firstly, the stellar objects discussed by Gokhroo and Mehra
\cite{36} are considered in which a specific form of the energy
density of the stellar system is assumed. We have observed that in
our case the effect of charge appears in Eqs.(\ref{54})-(\ref{56})
that describe the behavior of the system. Secondly, we have
considered stellar systems obeying the polytropic equation of state
and obtained a system of differential equations including the mass
equation, TOV equation and vanishing complexity condition in terms
of dimensionless variables. The solution of these equations for some
physical conditions provide a better understanding of the charged
stellar system with zero complexity factor. It is worthwhile to
mention here that all our results reduce to uncharged case ($q=0$)
\cite{19}.

\end{document}